\documentstyle[rawfonts,nassau,pslatex]{article}  
\frompage{000} \topage{000}                                              

\def\rightmark{Statistical Physics in a Finite Volume ...}
\def\leftmark{Sen Cheng}

\title{Statistical Physics in a Finite Volume with Absolute\\
  Conservation Laws} 
\authors{ 
  {Sen Cheng}\\[2.812mm]
  {\normalsize
    Department of Physics and Astronomy and\\
    National Superconducting Cyclotron Laboratory,\\
    Michigan State University, East Lansing, Michigan 48824-1321\\[0.2ex] 
    }
  }

\abstract{
  Recursion relations are used
  to exactly calculate the 
  partition function  of a
  canonical ensemble  in which  all additive
  charges as well as the total isospin are strictly conserved.
  The ensemble can consist of particles that 
  obey either classical or quantum statistics.
  Recursion relations are also employed to compute observables such as 
  multiplicity distributions and isospin fluctuations.
  }


\newcommand{\refEq}[1]{ Eq.~(\ref{#1})}

\begin{document}

\maketitle
\setcounter{page}{1}

\section{Introduction}\label{intro}

Results from a canonical ensemble that obeys exact conservation of
additive charges and isospin can be useful in a variety of applications.
For example, the unexpected isospin imbalances
of CENTAURO events \cite{Lattes:1980wk} could be
explained by employing a Gaussian thermal source and
restricting the pion emission to isoscalar pairs only.
The pion multiplicity distributions become broader as compared to those of
independent pion emission, thus resulting in larger fluctuations 
\cite{Pratt:1994tu,Pratt:1994cg}.
This finding could be further substantiated by replacing the 
Gaussian thermal source by a canonical ensemble and the ad-hoc assumption of 
isoscalar pairs emission by conservation of total isospin and its projection.

Another potential area of application is the modeling of relativistic heavy
ion collisions. Here, hydrodynamical models are used to describe the early
stages of the collision, when the density is so high that the particles behave
collectively as a fluid.
As the collision progresses the density falls, the 
system breaks up, and hydrodynamics is not applicable anymore.
Therefore, the break-up and later stages of the collision have to be described 
by other means such as, for example, microscopic transport models
\cite{Bass:2000ib}.
When switching to a microscopic transport model from a macroscopic
hydrodynamical prescription, particles must be formed from the fluid such that 
momentum and energy as well as isospin and all other charges are conserved.
Currently,  most schemes for generating particles are based on the grand
canonical ensemble.
However, exact conservation within each event and within each computational
cell is essential for studies on fluctuations and balance functions
\cite{Bass:2000az}.

Partition functions for canonical ensembles conserving charges
can be written quite easily, but they are very difficult
to compute for large systems because the number of partitions that have to be
summed over increases rapidly
\cite{Cleymans:1997ib}.
This difficulty can be overcome by rewriting the partition function
as a recursion relation.
These recursion relations were applied  to ensembles of
classical particles in \cite{Pratt:1999ht}  and  to Fermi systems in
\cite{Pratt:1999ns}.

In this work we add total isospin conservation to the
canonical ensemble and derive the appropriate recursion relations.
Furthermore, isospin fluctuations in the canonical ensembles, which, for
example, are relevant for CENTAURO event studies, are calculated.
A Monte Carlo algorithm that employs these recursion relations to generate
particles is part of ongoing work and will be presented in the future.
As our focus is the development of the computational technique, we
show sample calculations for simple systems
rather than comparisons to experimental results.

\section{Recursion Relations with Classical Statistics}
\label{sec:class}

\subsection{Partition Functions}
\label{sec:class:partition}

The partition function for a canonical ensemble of  $A$ particles conserving a
charge $Q$, where $Q$ could be a vector containing several charges, is defined 
as
\begin{equation}
  \label{eq:pf_def}
  Z_{A,Q} = \sum_\alpha \langle \alpha | e^{-\beta H} | \alpha \rangle ,
\end{equation}
where $\alpha$ labels some quantum state with the appropriate quantum numbers,
$\beta$ is the inverse temperature,
and $H$ is the Hamiltonian of the system.
Using the single particle partition function
\begin{equation}
  \label{eq:omega_k}
  \omega_k= \sum_i \exp(-\epsilon^{(k)}_i/T),
\end{equation} 
which is simply a sum of Boltzmann weights over all available states
$i$, the partition function can be written in a more tractable form
as a sum over partitions subject to particle number and charge conservation
\begin{equation}
  \label{eq:pf_sum_partitions}
  Z_{A,Q} = \sum_{\langle \sum \nu_k a_k = A; \sum \nu_k q_k = Q\rangle}
  \prod_{k=1}^{N} \frac{\omega_k^{\nu_k}} {\nu_k!}.
\end{equation}
$N$ is the number of particle types,
$\nu_k$ is the number of particles of type $k$, and
$q_k$ is the charge of one particle.
The particle number $a_k$ indicates how many times a particle contributes to
the main conserved quantity,
e.g., 
in the case of multi-fragmentation, $a_k$ would be the mass number of the
nuclei.

The computational challenge of \refEq{eq:pf_sum_partitions} is hidden in
the number of partitions to be summed over. Assuming $a_k=1$ for all $k$,
the number of partitions is 
\begin{equation}
  J(A,N)= \frac{(A+1)!} {(A+2-N)!},
\end{equation}
which is prohibitive to compute for large $A$ or $N$.
However, summing over all partitions can be bypassed when using recursion
relations as derived in \cite{Pratt:1999ht}
\begin{equation}
  \label{eq:pf_recursive}
  Z_{A,Q} = \frac1A \sum_j a_j \omega_j Z_{A-a_j, Q-q_j}.
\end{equation}
If intermediate values are stored, \refEq{eq:pf_recursive} requires a linear
runtime in both, $A$ and $N$.

To conserve isospin as well, a sum over all possible isospin configurations
for a given partition, $\{\nu_k\}$, 
and isospin weights $W\left(I,M|\{\nu_k\}\right)$ have to be included in the partition function
\begin{equation}
  \Omega_{A,I,M,Q} = 
  \sum_{\langle \sum \nu_k a_k = A; \sum \nu_k q_k = Q\rangle}
  \sum_{\{\nu_k\}} W\left(I,M|\{\nu_k\}\right)
  \prod_{k=1}^{N} \frac{\omega_k^{\nu_k}} {\nu_k!}.
\end{equation}
This partition function can be converted to a recursion relation by
inserting $\frac 1A {\sum_{j=1}^{N} a_j \nu_j} =1$ 
\begin{equation}
  \Omega_{A,I,M,Q} = 
  \frac 1A
  \sum_{j=1}^{N} a_j \omega_j
  \sum_{\langle \sum \nu_k a_k = A; \sum \nu_k q_k = Q\rangle}
  \sum_{\{\nu_k\}} W\left(I,M|\{\nu_k\}\right)
  \frac {\omega_j^{\nu_j-1}} {(\nu_j -1)!}
  \prod_{k\neq j} \frac{\omega_k^{\nu_k}} {\nu_k!}.
\end{equation}
The summation indices can be switched
\begin{equation}
  \nu'_k = \left\{
    \begin{array}{ll}
      \nu_k     & , k\neq j \\
      \nu_k -1  & , k=j
    \end{array}
  \right.
\end{equation}
if the sum over isospin configurations is split:
\begin{equation}
  \sum_{\{\nu_k\}} W\left(I,M|\{\nu_k\}\right)
  =
  \sum_{I'=|M-m_j|}^{I_j+I} \langle I_j m_j; I', M-m_j| IM \rangle ^2
  \sum_{\{\nu'_k\}}  W\left(I',M|\{\nu'_k\}\right).
\end{equation}
Finally, the recursion relation with conserved total isospin turns out to be
\begin{equation}
  \Omega_{A,I,M,Q} = \frac{1}{A} \sum_j a_j\omega_j 
  \sum_{I'=|M-m_j|}^{I+I_j} \langle I_jm_j; I',M-m_j | IM\rangle^2 
  \Omega_{A-a_j,I',M-m_j,Q-q_j} .
\end{equation}

\subsection{Multiplicity Distributions}
\label{sec:class:multi}

We derive a recursion relation for the multiplicity distribution 
of a set  $B$ of particle types.
The probability of observing $n$ particles of a type contained in the set $B$
can be expressed in terms of factorial moments
\cite{Pratt:1999ht} 
\begin{equation}
  P_{B,A,I,M}(n)=
  \sum_{l\geq n} F_{B,A,I,M}(l) \frac{(-1)^{l-n}} {(l-n)! n!} ,
\end{equation}
for which recursion relations can be found more easily. Following the
procedure in the previous subsection the recursion relation for the factorial
moment is found to be
\begin{equation}
  \label{eq:factorial_mom}
  F_{B,A,I,M}(n) = 
  \sum_{k\in B} \omega_k 
  \sum_{I'} \langle I_km_k; I',M-m_k | IM\rangle^2 
  \frac{\Omega_{A-a_k,I',M-m_k}}{\Omega_{A,I,M}} 
  F_{B,A-a_k,I',M-m_k}(n-1),
\end{equation}
which leads to the recursion relation for the multiplicity distribution
\begin{equation}
  \label{eq:multi}
  P_{B,A,I,M}(n)=
  \frac1n  \sum_{k\in B} \omega_k 
  \sum_{I'} \langle I_km_k; I',M-m_k | IM\rangle^2 
  \frac{\Omega_{A-a_k,I',M-m_k}}{\Omega_{A,I,M}} 
  P_{B,A-a_k,I',M-m_k}(n-1).
\end{equation}

\begin{figure}[htb]
  \vspace*{-4mm}
  \insertplot{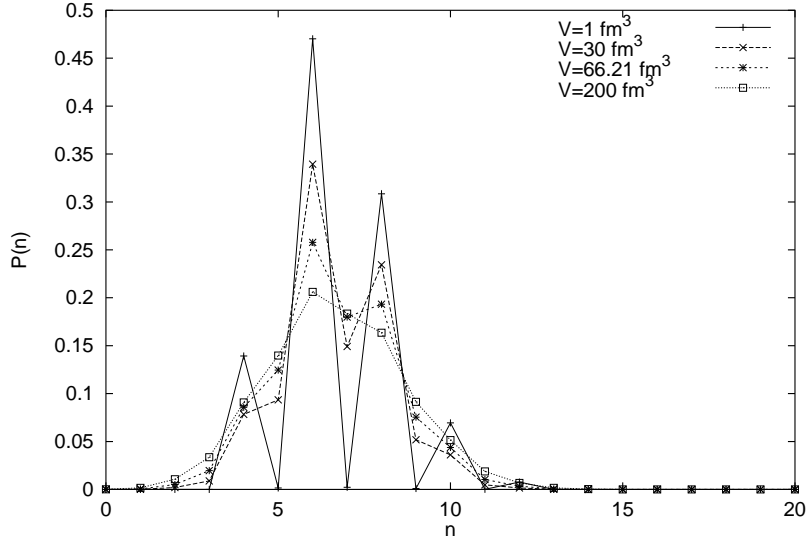}
  \vspace*{-8mm}
  \caption[]{Multiplicity distribution of neutral pions for several volumes.
    The calculation
    includes all pseudo-scalar and vector mesons and their weak decays.
    The system was constrained to an isospin singlet  at a temperature of 
    $170$~MeV.
    The total number of pions was a fixed $A=20$.
    }
  \label{fig:multi}
\end{figure}

To test \refEq{eq:multi} we implemented a program that included 
the pseudo-scalar and vector mesons as well as their decays. 
Figure \ref{fig:multi} shows the multiplicity distribution for neutral pions:
the system was constrained to be in an isospin singlet at a temperature of
$170$~MeV and had to produce a total of 20 pions.
For this and all following sample calculations 
we need to make an assumption about the energy states that are available to the
particles and that have to be
summed over in the single particle partition function of
\refEq{eq:omega_k}.
We required that the linear momentum $k$ in one dimension
satisfy the condition
\begin{equation}
  kR= \pi \left( n+ \frac 12 \right),
\end{equation}
where $n= 0, 1, 2, \ldots$ and $R$ is the size of the system. 
The half integers on the RHS, instead of the more usual integers,  were
inspired by the MIT bag model and lead to a reduction of zero-point
surface energy effects. This becomes important for systems that are confined
to a small volume. By assuming that our model systems are confined to a cube
of volume $V$, we obtain the energy states
\begin{equation}
  \epsilon_{n,m,l}= \sqrt{\frac{\pi^2}{R^2} \left[ 
      (n+1/2)^2 + (m+1/2)^2 + (l+ 1/2)^2 \right]
    + M^2
    },
\end{equation}
where $n,m,l= 0, 1, 2, \ldots$, $M$ is the mass of the particle, and 
$R= V^{1/3}$.

The even-odd asymmetry in Fig.\ \ref{fig:multi} for small volumes that
vanishes for larger volumes can be understood in the following way. At small
volumes and low temperatures the system minimizes the number, not the mass, of
particles produced; it prefers to produce heavy particles that decay into
multiple pions rather than producing pions or kaons. These heavy particles
($\omega$, $\eta$, $\eta'$) decay into an odd number of pions (and also into
an odd number of neutral pions). Because the number of heavy particles has to
be even to make 20 pions, as we imposed, the number of neutral pions is
strongly favored to be even. At higher volumes or temperatures low mass
particles dominate and the even-odd asymmetry vanishes. 

\begin{figure}[htb]
  \vspace*{-4mm}
  \insertplot{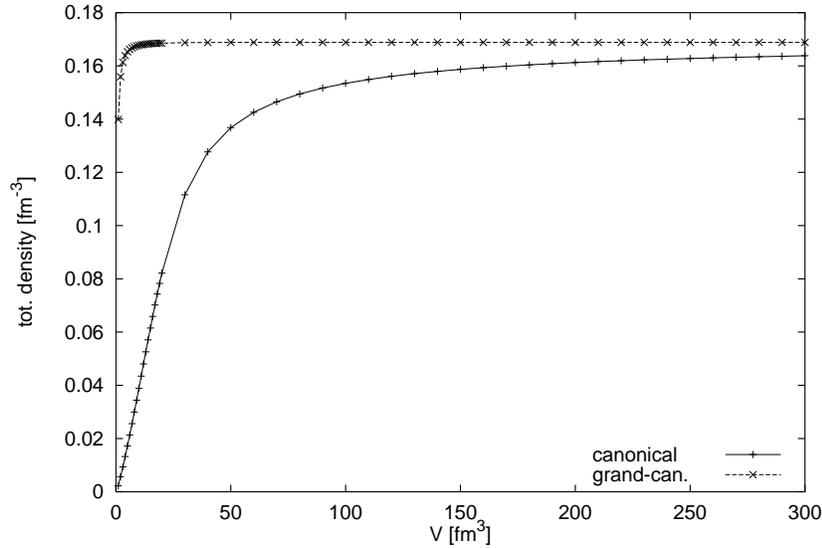}
  \vspace*{-8mm}
  \caption[]{
    Total number density of a system of only pions in an isosinglet at a
    temperature of $170$~MeV. Shown are calculations for a grand-canonical
    and a canonical ensemble with classical statistics.
    }
  \label{fig:density}
\end{figure}

The factorial moment  from \refEq{eq:factorial_mom} corresponds to the density,
when $n=1$. For a system of pions in an isosinglet, 
particle production is significantly suppressed at low volume, but converges
towards the grand-canonical result at large volumes  (Fig.\ \ref{fig:density}).

\section{Recursion Relations with Quantum Statistics}
\label{sec:qm}


For a canonical ensemble of quantum particles  with charge conservation the
recursion relation for the partition function was given in \cite{Pratt:1999ns}:
\begin{equation}
  Z_{A,Q} =
  \frac 1A \sum_k \sum_{n=1}^\infty C_n^{(k)} Z_{A-n, Q-  q_k  } ,
\end{equation}
where
$C_n^{(k)}=(\pm1)^{n-1} \sum_i \exp(-n\epsilon_i^{(k)}/T)$.
To derive a partition function $ \Omega_{A,I,M,Q}$ 
that conserves total isospin as well, we first
observe that, if the Hamiltonian of the system is independent of $M$, then so
is the partition function. Then the partition function conserving total
isospin can be related to one that conserves only the projection 
$  Z_{A,M,Q}= \sum_{I\geq M} \Omega_{A,I,M,Q}$.
Therefore
\begin{equation}
  \Omega_{A,I,M,Q}= Z_{A,M,Q}-Z_{A,M+1,Q}.
\end{equation}


To calculate densities  the 2-point function is needed
\begin{equation}
  \langle a_i^\dagger a_j\rangle= \frac{\delta_{ij}}{Z_A} \sum_n
  (-1)^{n-1}\exp(-n\epsilon_i/T) Z_{A-n},
\end{equation}
which, however, can only be used when conserving additive quantities, but not
isospin. 
The 4-point function serves to calculate higher moments
\begin{equation}
  \langle a_i^\dagger a_j^\dagger a_k a_l\rangle =
  \frac{1}{Z_A} (\delta_{il}\delta_{jk}\pm \delta_{ik}\delta_{jl})
  \sum_{n_i,n_j} (-1)^{n_i+n_j}
  \exp(-n_i\epsilon_i/T-n_j\epsilon_j/T) 
  Z_{A-n_i-n_j} .
\end{equation}

\section{Isospin Fluctuations}
\label{sec:fluc}
For classical particles the isospin fluctuation
\begin{equation}
  \label{eq:fluc_def}
  \sigma= \sum_{I=0} \langle I | (n_+ + n_- - 2n_0)^2 | I \rangle,
\end{equation}
where $n_+$, $n_-$, $n_0$ are the densities of the respective pions, 
can be readily computed with the recursion relations derived above.
However, for quantum particles we cannot compute  the density, let alone the
higher moments,  when conserving total isospin. Therefore, we need to rewrite
\refEq{eq:fluc_def} in terms of isospin projection states.

The operator in \refEq{eq:fluc_def}, itself a product of spherical tensor
components, can be written as a decomposition into other 
spherical tensor components
$A_{JM}$
\begin{equation}
  (n_+ + n_- - 2n_0)^2
  = 6 T_{20}T_{20}
  = 6 \sum_{J,M} \langle 20; 20 | JM \rangle A_{JM}.
\end{equation}
By the Wigner-Eckart theorem only $A_{00}$ contributes when contracted between 
isosinglet states,
$\sigma= 6 \sum_{I=0} \langle I | A_{00} | I \rangle$,
which, after some algebra, becomes
\begin{equation}
  \sigma= \frac65 \left(\sum_{M=0}-\sum_{M=1} \right) 
  \left\langle M |
    \frac32 n_+ + \frac32 n_- + n_0 + \frac16\left(n_+ + n_- - 2n_0\right)^2 | M 
  \right\rangle .
\end{equation}

\begin{figure}[htb]
  \vspace*{-4mm}
  \insertplot{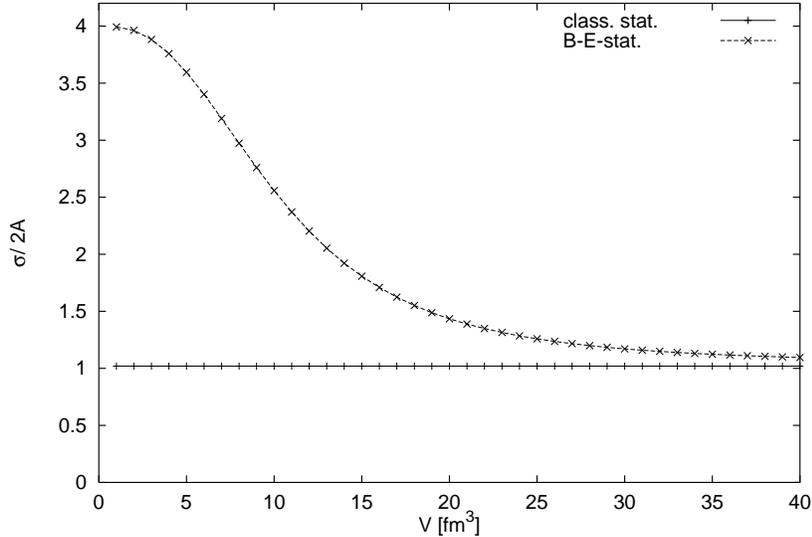}
  \vspace*{-8mm}
  \caption[]{
    Isospin fluctuations as a function of volume for a system of pions at a
    temperature of $170$~MeV with a fixed number of pions, $A=20$.
    Shown are calculations with classical and Bose-Einstein statistics. The
    isospin fluctuations are divided by $2A$, the expected isospin fluctuation
    for a gas of independent pions.
    }
  \label{fig:fluc_V}
  \vspace*{-5mm}
\end{figure}

\begin{figure}[htb]
  \vspace*{-4mm}
  \insertplot{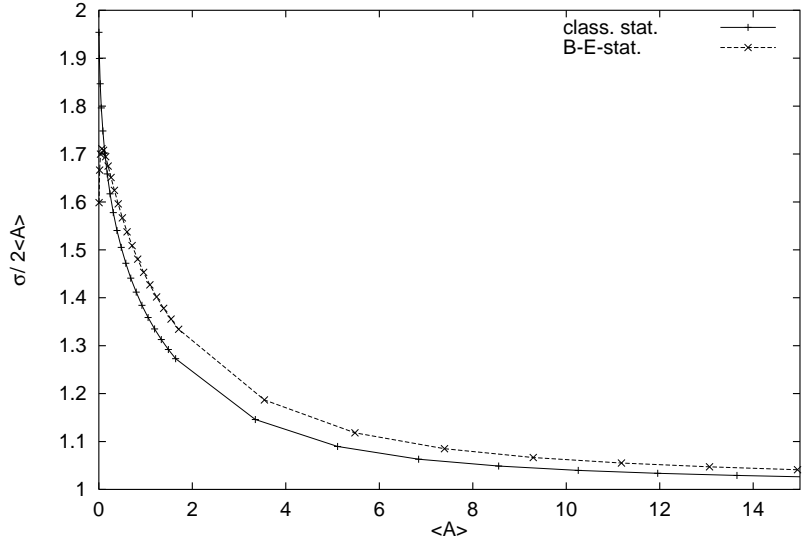}
  \vspace*{-8mm}
  \caption[]{
    Isospin fluctuations as a function average pion number
    for a system of pions at a temperature of $170$~MeV.
    Shown are calculations with classical and Bose-Einstein statistics. The
    isospin fluctuations are divided by $2\langle A\rangle$, the expected isospin fluctuation
    for a gas of independent pions.
    }
  \label{fig:fluc_A}
  \vspace*{-5mm}
\end{figure}

In Figs.\ \ref{fig:fluc_V} and \ref{fig:fluc_A} we compare classical to
Bose-Einstein statistics in
calculations of isospin fluctuations for a system of pions at $T=170$~MeV.
The fluctuations
are divided by $2A$, the expected fluctuation for an independent pion gas, to
highlight the difference. When the system is fixed to contain 20  pions
(Fig.\ \ref{fig:fluc_V}), the isospin fluctuation for classical particles
does not depend on the volume, as the relative composition of the classical
pion gas does not. Bose-Einstein particles, however, prefer to be all in
the same state, when the volume becomes small, thus enhancing the isospin
fluctuations.
When the number of pions is not fixed, the isospin
fluctuations converge towards the value for an independent pion gas as the
system increases in size (Fig.\ \ref{fig:fluc_A}). 

\section{Conclusions}
\label{concl}

Recursion relations have been derived for a canonical ensemble  in which  all
additive charges as well as the total isospin are strictly conserved. 
In the case of classical statistics we derived recursion relations
for the partition function and multiplicity distribution. For quantum
statistics we obtained a recursion relation for  the
partition function and density. Work is under way to 
derive the recursion relation for the multiplicity distribution in the latter
case.

These recursion relations can be applied to an arbitrary number of particle
types and conserved additive charges. They can be computed correctly within
the numerical precision of the computer in a reasonable runtime as our sample
calculations show. 
We can now move forward and apply this technique to physical problems
and in Monte Carlo algorithms that generate particles with exact
conservation of additive charges and isospin.
We also derived a method to quickly calculate isospin fluctuations, which makes
use of  the recursion relations  we obtained. 

\section*{Acknowledgements}
This work was supported by the National Science Foundation, Grant No.
PHY-00-70818.


\vfill\eject
\end{document}